\documentstyle[prl,preprint,aps]{revtex}
\begin{document}
\preprint{UCSBTH-93-24}
\draft

\title{\Large{AC Stark Effects and Harmonic Generation in Periodic Potentials}}

\author{Martin Holthaus}
\address{Department of Physics, Center for Nonlinear Sciences,
        and Center for Free--Electron Laser Studies,
        University of California, Santa Barbara, CA 93106}

\author{Daniel W.~Hone}
\address{Department of Physics and Center for Quantized Electronic Structures,
        University of California, Santa Barbara, CA 93106}

\date{\today}
\maketitle

\begin{abstract}
The ac Stark effect can shift initially nonresonant minibands in
semiconductor superlattices into multiphoton resonances.  This effect
can result in strongly enhanced generation of a particular desired harmonic
of the driving laser frequency, at isolated values of the amplitude.
\end{abstract}

\pacs{PACS numbers: 73.20.Dx, 73.40.Gk, 42.65.Ky}

\narrowtext
The spectral structure of an atom can be substantially modified by an
intense laser field.  In particular, at sufficiently high field strengths
the ac Stark effect can bring initially nonresonant levels into multiphoton
resonances, resulting, {\it e.g.}, in an enhancement of the
photoionization signal \cite{FRE}.  Recent experiments \cite{STO,VRI}
have confirmed that ac Stark effects are of central importance for a
detailed understanding of the ionization dynamics \cite{SLF,AGO}.

If we deal with a periodic lattice, instead of an atom, atomic levels
broaden into energy bands.  The question then arises whether a laser field
can similarly be used to modify the effective band structure in a
controllable way, so as to alter physical behavior.  This question is
of particular interest and importance for electron minibands in
semiconductor superlattices \cite{BAS}, where the miniband structure itself
 can be engineered, within wide ranges, by the design process \cite{SUP}.
Also, the relevant spatial size and energy scales for these structures are
such that the interesting range of electromagnetic frequencies is in the
far infrared, and there are now sources available which can probe them
\cite{GUI} with nonperturbatively strong laser fields in this spectral
range.  In this Letter we will show that ac Stark effects play a major role
 in these systems, as well.  We  first outline the necessary theoretical
formalism, and we then demonstrate by simple numerical examples the
practicality of constructing devices which make use of field-induced
multiphoton resonances in superlattices for the selectively enhanced
generation of particular harmonics of the driving field.

We consider a particle with effective mass $m^*$ moving in a one-dimensional
periodic potential $V(x)$ with lattice constant $a$.  When this system
interacts with an external, spatially homogeneous, electric field ${\cal
E}(t)={\cal E}_0\sin(\omega t)$, the Hamiltonian is given by
\begin{equation}
{\cal H}(x,t) = \left[p-eA(t)\right]^2/2m^* + V(x) \;\;\; , \label{HXT}
\end{equation}
where $A(t)$ is the electromagnetic vector potential: ${\cal E}(t) =
-dA(t)/dt$.  Because this Hamiltonian is periodic in $x$, the wave vector
$k$ remains a good quantum number in the presence of the external field.
When transitions between different bands can be completely neglected, so
that only the dynamics within a single band need be considered, then it is
straightforward \cite{HOU,KRI,QWS} to solve the time-dependent Schr\"odinger
equation: if $\varphi_k(x) = \exp(ikx)v_k(x)$ are the Bloch wave eigenstates
for the unperturbed lattice, with energy $E(k)$, then the states
\begin{equation}
\psi_k(x,t) = e^{ikx}v_{q(t)}(x)\exp\left\{-i\int_0^t \!d\tau\,
E[q(\tau)]\right\}
\label{PSI}
\end{equation}
approximately satisfy the full time dependent equation $i\partial_t\psi_k(x,t)
= {\cal H}(x,t)\psi_k(x,t)$, provided that $q(t)$, which labels the periodic
part of the Bloch functions $v_{q(t)}(x)$, is given by $q(t) = k-eA(t)$.
Thus, this function $q(t)$ obeys \cite{ZIM,KIT} the classical equation of
motion,
\begin{equation}
\dot q(t) = e{\cal E}(t). \label{CLA}
\end{equation}
This ``acceleration theorem'' is usually the starting point \cite{BAS} for a
discussion of carrier motion in semiconductor superlattices in high-frequency
electric fields.

However, there is another, more general formulation of this problem.  The
Hamiltonian (\ref{HXT}) is periodic not only in space, but also in time,
with period $T=2\pi/\omega$.  The Floquet theorem then guarantees \cite{ZPB}
the existence of a complete set of solutions to the time dependent
Schr\"odinger equation of the form
\begin{equation}
\psi(x,t) = \exp\left\{i\left[kx-\varepsilon(k)t\right]\right\}w_k(x,t) \;\; ,
\label{WFN}
\end{equation}
with $w_k(x,t)=w_k(x+a,t)=w_k(x,t+T)$, and with ``quasienergies''
$\varepsilon(k)$, which specify the effect of time translation by a full
 period $T$.  Thus, in the presence of the external field the energy $E(k)$
is effectively replaced by the quasienergy  $\varepsilon(k)$ in
characterizing the time evolution of the state labeled by quasimomentum $k$.
Defining  the Floquet functions $u_k(x,t)\equiv \exp(ikx)w_k(x,t)$, we can
find the quasienergies from the eigenvalue equations
\begin{equation}
\left[{\cal H}(x,t)-i\partial_t\right]u_k(x,t) = \varepsilon(k)u_k(x,t),
\label{EIG}
\end{equation}
with appropriate boundary conditions in space and time.  Within the single
band approximation quasienergies are easily determined from the wave
functions~(\ref{PSI}): $v_{q(t)}$ is already periodic in space and in time,
and the phase grows, on average, linearly with time.  The quasienergy  is
then given \cite{QWS} by the average growth rate,
\begin{equation}
\varepsilon(k) = {1\over T}\int_0^T \!d\tau\, E\left[q(\tau)\right].
\label{AVE}
\end{equation}
For example, the standard cosine dispersion for a tight binding band of
width $W$, $E(k)=(W/2)\cos(ka)$, yields quasienergies
$\varepsilon(k)=(W/2)J_0(e{\cal E}_0a/\omega)\cos(ka)$, with $J_0$ the zero
order Bessel function.

It is crucial to recognize that the transition from the classical equation
of motion~(\ref{CLA}) to the quantum eigenvalue equation~(\ref{EIG}) is not
merely a reformulation.  The eigenvalue equation is significantly more
general; it provides a framework for attacking problems which lie outside
the scope of a semiclassical treatment.  For example, quasienergies remain
well defined if $k$ is no longer an exact quantum number, either because of
finite lattice size or because of lattice imperfections.  If the departures
from spatial periodicity are local, {\it e.g.}, then~(\ref{EIG}) is the
starting point for an extension \cite{NEW} of standard Green's function
treatments of spatially local defects to systems subjected to strong time
periodic fields. This reduces the problem to finite quadratures in terms of
the solutions to the defect free problem.  Most importantly, the existence of
dispersion relations $\varepsilon_n(k)$ (where $n$  is a band index) relies
only on the lattice spatial periodicity and the temporal periodicity of
${\cal E}(t)$.  Therefore, these quasienergies can still be defined
rigorously even when laser-induced transitions between unperturbed energy
bands are no longer negligible, so that (\ref{CLA}), and
therefore~(\ref{AVE}), become invalid.  Knowledge of the $\varepsilon_n(k)$
implies a nonperturbative understanding of inter(mini)band effects.  It is
this fact, and its consequences for possible experiments, that we will now
explore in detail.

Fig.~\ref{fig1} shows the lowest three quasienergy bands, calculated
numerically, for a model potential $V(x)$ consisting of 20 square wells of
width 330 \AA{} which are separated by rectangular barriers of width 40 \AA\
and height 0.3~eV.  The effective particle mass is chosen as that of an
electron in the conduction band of GaAs, $m^*=0.067 m$ (with $m$ the bare
electron mass), and the external frequency as $\omega = 3.0$ meV, as an
approximate model of a GaAs/AlGaAs superlattice in a far-infrared laser field
(linearly polarized in the growth direction).  If $\varepsilon$ is a
quasienergy, then so is $\varepsilon + m\omega$ for any integer $m$.
Therefore, it suffices to consider only a single quasienergy ``Brillouin
zone'', a range of quasienergies of width $\omega$.  The bands in
Fig.~\ref{fig1} are labelled such that ($n$, $-m$) denotes the representative
of the $n$th quasienergy band which is shifted down in energy by $m\omega$
from that representative which is connected to the unperturbed $n$th energy
band for vanishing laser field strength.

The first and second energy bands in the unperturbed model are separated by a
gap of 12.93 meV, which is more than four times the photon energy $\omega$.
With increasing field amplitude ${\cal E}_0$ the bands in Fig.~\ref{fig1}
oscillate in width and almost collapse (apart from an almost degenerate pair
of edge states on top of each band) at values of $e{\cal E}_0a/\omega$ equal
to a zero of the Bessel function $J_0$.  This is the behavior characteristic
also of isolated bands.  But, in addition, interaction between different bands
leads to an ac Stark shift: the lowest band (labelled $n=1$) is shifted down
when the field becomes stronger, whereas the first excited band ($n=2$) is
shifted up.  Then there necessarily exists a critical field strength where
the gap between the bands approaches $5\omega$, so that we have a
field-induced 5-photon resonance.  In a quasienergy plot reduced to a single
``Brillouin zone'' such a resonance manifests itself as an avoided crossing
of quasienergy bands, as shown in Fig.~\ref{fig1} for the representatives
(1,0) and (2,-5).

At the field strength of an avoided crossing, two bands are strongly coupled.
In a description based on the original energy bands one would at least have
to include strong multiphoton transitions between them by higher order
perturbation theory. A nonperturbative description based on quasienergy states
is simpler, and more general, since the eigenvalue equation~(\ref{EIG}) fully
incorporates the effect of the ac field.  The Floquet states are time dependent
linear combinations of the unperturbed band states, with maximal interband
mixing at an avoided crossing \cite{WIN}, so that the Floquet picture {\it
automatically} includes the multiphoton transitions between different
unperturbed bands.

A field-induced avoided crossing of quasienergy bands can have observable
consequences.  Strong coupling between different bands, which occurs only near
particular values of the amplitude ${\cal E}_0$, implies that the
characteristics of the harmonics generated by the periodically driven lattice
should alter dramatically at those values.  To demonstrate this we choose an
arbitrary single Floquet state $\psi(x,t)$ from the lowest band and Fourier
decompose its dipole expectation value,
\begin{equation}
\langle\psi(t)|x|\psi(t)\rangle = \sum_nx_n\exp(-in\omega t). \label{DIP}
\end{equation}
Fig.~\ref{fig2} shows the Fourier coefficients $|x_n|$ for field strengths
${\cal E}_0$ below (8500 V/cm), at (9250 V/cm), and above (10~000 V/cm) the
avoided crossing.  Where the 5-photon resonance occurs, the fifth harmonic is
strongly enhanced.  The response at this frequency is even stronger than it
is at the fundamental frequency $\omega$.  The most general solution to the
Schr\"odinger equation is a superposition of all Floquet states, and
interferences can lead to the occurrence of additional frequencies in the
dipole moment~(\ref{DIP}).  However, the principal feature --- strong
enhancement of the $m$th harmonic because two bands are ac Stark shifted into
an $m$-photon resonance --- is a general result.  This effect can also be
found without recourse to Floquet theory  by numerically analyzing the Fourier
content of solutions to the Schr\"odinger equation at different values of the
amplitude ${\cal E}_0$.  The advantage of the Floquet method is that one does
not have to search blindly for large effects as a function of ${\cal E}_0$,
but can identify multiphoton resonances immediately from a quasienergy
diagram.  Thus, we now can formulate a general rule: a possible device for
efficient harmonic generation by electrons in spatially periodic potentials
and strong ac fields should be operated at a critical point of the
quasienergy-quasimomentum dispersion relation --- {\it i.e.}, at an avoided
crossing of quasienergy bands.

Application of this principle to semiconductor superlattices is particularly
interesting, because these artificial lattices can be engineered so as to
optimize the effect. Consider a simple one-dimensional tight binding
Hamiltonian, which provides a good description of tunneling between different
superlattice wells,
\begin{equation}
{\cal H}_{tb} = E_0\sum_\ell|\ell\rangle\langle\ell| + {W\over 4}\sum_\ell
\left[|\ell +1\rangle\langle\ell| + |\ell\rangle\langle\ell +1|\right],
\label{TBH}
\end{equation}
where $\{|\ell\rangle\}$ is a set of Wannier states localized in the
individual wells.  This Hamiltonian contains two parameters, the on-site
energy $E_0$ and the hopping strength $W/4$.  If we now start with a system
described by~(\ref{TBH}), and then modify either the on-site energy (the well
width) of every second site, or every second hopping integral (barrier width),
then the unit cell of the periodically repeated structure consists of two
wells, and the original bands each split into two. In this way it is possible
to create lattices where two bands (or more, if the number of wells within a
unit cell is increased further) are grouped arbitrarily close in energy, so
that inter(mini)band effects play a dominant role even for small applied field
amplitudes.

To illustrate this we now use a model potential $V(x)$ of 50 square wells each
of width 90 \AA .  The separating barriers are 0.3 eV high, and their widths
alternate between 40 and 60 \AA .  As before, the effective mass $m^*$ is
0.067 in units of the bare electron mass.  The lowest doublet for this
dimerized system consists of two very narrow energy bands which are separated
by a gap of 1.7 meV; this doublet is in turn separated by the much larger gap
of approximately 100 meV from the first excited doublet.  Fig.~\ref{fig3} shows
the quasienergy bands that originate from the lowest doublet, for an applied
field of frequency $\omega = 1.0$ meV.  In the two original unperturbed bands
(for ${\cal E}_0=0$)  corresponding states (those with the same wave vector
$k$) are separated by an energy less than 3 meV.  With increasing field
strength the ac Stark effect pushes these two bands further apart, resulting
first in a 3-photon resonance ($3\omega =3$ meV) and the corresponding avoided
quasienergy band crossing near ${\cal E}_0\approx 2200$ V/cm (see
Fig.~\ref{fig3}), and then a 5-photon resonance and the related avoided
crossing near ${\cal E}_0\approx 4900$ V/cm. The corresponding enhancement of
the fifth harmonic generation in the latter case is exhibited in
Fig.~\ref{fig4}, which shows the Fourier amplitudes of the dipole moment
corresponding to one of the states at this avoided crossing. These field
strengths and superlattice parameters are well within practical experimental
ranges.  In general, the potential efficiency of harmonic generation associated
with an avoided crossing is determined by its sharpness (the curvature of the
quasienergy as a function of field amplitude at the avoided crossing; the
sharper, the more efficient), which can be controlled by varying system
parameters.

More than 20 years ago Tsu and Esaki \cite{TSU} realized that the nonlinear
optical response of conduction electrons in superlattices might find important
device applications.  The nonlinearity they considered explicitly was
associated with the nonparabolicity of a single isolated superlattice miniband.
 We have shown that nonlinearities can be strongly enhanced by {\it
inter}miniband effects which are induced by strong ac electric fields.  Even
in the presence of these time periodic fields the quantum behavior is simply
and usefully described in terms of a few numbers --- quasienergy and (for
finite systems, approximate) quasimomentum.  Using such a description we have
demonstrated the possibility of employing the ac Stark effect to tune
minibands into multiphoton resonances, and that this implies the possibility
of selectively enhanced generation of a particular harmonic.  The fabrication
of superlattices has reached  a state of great perfection~\cite{SUP}, and the
first experiments \cite{GUI} with superlattices in strong far-infrared laser
fields have been successfully carried out, so it should be possible now to
observe and exploit this predicted effect.  More generally, the further
investigation of superlattices in strong ac fields seems highly promising in
terms of advances in both pure and applied physics.

We are grateful to S.J. Allen for helpful discussions.  This work was
supported in part by the Alexander von Humboldt Foundation and by the Office
of Naval Research under Grant No. N00014-92-J-1452.

\begin{figure}
\caption{Lowest 3 quasienergy bands for 20 square wells of width 330 \AA,
separated by barriers of width 40 \AA\ and height 300 meV.  The particle mass
is $m^*=0.067m$, and the ac frequency $\omega=3.0$ meV. The inset shows a
magnified view of the avoided crossing of the lowest two bands.}
\label{fig1}
\end{figure}

\begin{figure}
\caption{Fourier coefficients $|x_n|$ of the dipole expectation value  for an
arbitrary state in the lowest band of Fig.~\protect\ref{fig1}, at field
strengths ${\cal E}_0$ of (a) 8500 V/cm, (b) 9250 V/cm, and (c) 10~000 V/cm.}
\label{fig2}
\end{figure}

\begin{figure}
\caption{Quasienergy bands from the lowest doublet, corresponding to electrons
of effective mass $m^*=0.067m$ in 50 square wells of width 90 \AA, separated
by square barriers of height 300 meV and widths alternating between 40 \AA\
and 60 \AA; the ac frequency is $\omega=1.0$ meV. Two Brillouin zones are
shown.}
\label{fig3}
\end{figure}

\begin{figure}
\caption{Dipole expectation value Fourier coefficients for an arbitrary state
at the  avoided crossing near 4900 V/cm in Fig.~\protect\ref{fig3}.}
\label{fig4}
\end{figure}

\end{document}